\documentclass[aps,prl,twocolumn,floats,amsmath,amssymb,showpacs,floatfix]{revtex4}
\usepackage{graphicx}
\usepackage{dcolumn}
\usepackage{psfrag}
\usepackage{float}
\newcommand{\ve}[1]{\ensuremath{\mbox{\boldmath$#1$}}}
\newcommand{\ma}[1]{\ensuremath{\mathbb{#1}}}

\newcommand{\ku}{\ensuremath{\mbox{Ku}}}
\newcommand{\st}{\ensuremath{\mbox{St}}}
\newcommand{\tr}{\ensuremath{\mbox{Tr}}}

\newcommand{\Ordo}{\ensuremath{O}}

\begin{document}
\title{Ergodic and non-ergodic clustering of inertial particles}
\author{K. Gustavsson and  B. Mehlig}
\affiliation{Department of Physics, Gothenburg University, 41296
Gothenburg, Sweden}

\begin{abstract}
We compute the fractal dimension of clusters of inertial
particles in mixing flows at finite values of Kubo ($\ku$) and 
Stokes ($\st$) numbers,
by a new series expansion in $\ku$. At small $\st$,
the theory includes clustering by Maxey's non-ergodic \lq centrifuge' effect.
In the limit of $\st\rightarrow \infty$ and $\ku \rightarrow 0$
(so that $\ku^2 \st$ remains finite) 
it explains clustering in terms of ergodic \lq multiplicative amplification'.
In this limit, the theory is consistent with the asymptotic
perturbation series in
[Duncan {\em et al.}, Phys. Rev. Lett. {\bf 95} (2005) 240602].
The new theory allows to analyse how 
the two clustering mechanisms compete
at finite values of $\st$ and $\ku$. 
For particles suspended in two-dimensional  random Gaussian incompressible flows,
the theory yields excellent results for $\ku < 0.2$
for arbitrary values of $\st$; the ergodic mechanism is found to contribute
significantly unless $\st$ is very small. For higher values of $\ku$
the new series is likely to require resummation. But numerical simulations
show that for $\ku\sim \st \sim 1$ too, 
ergodic \lq multiplicative amplification' makes 
a substantial contribution to the observed clustering.
\end{abstract}

\maketitle

\begin{figure}[t]
\centerline{
\includegraphics[width=7.5cm]{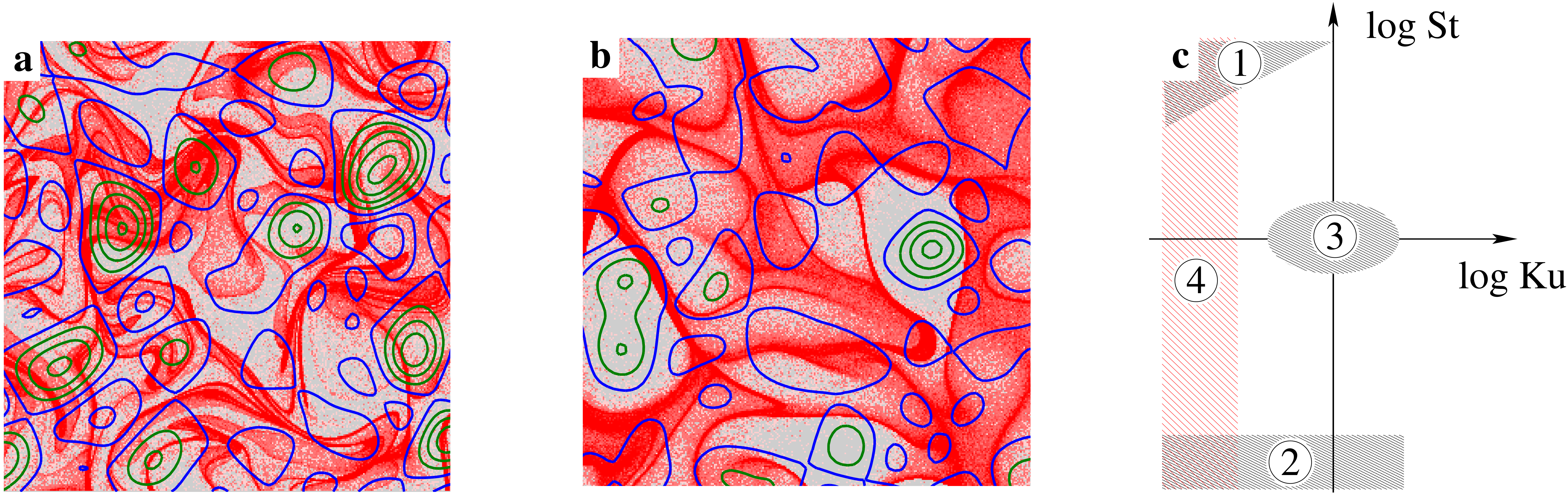}}
\caption{\label{fig:1}
{\bf a} Clustering of particles in
 a two-dimensional
incompressible flow $\ve u(\ve r,t)=\ve\nabla\wedge\psi(\ve r,t) {\bf e}_3$.
Here ${\bf e}_3$ is the unit vector $\perp$  to the plane.
The Gaussian  random
function $\psi(\ve r,t)$ satisfies $\langle \psi\rangle = 0$ and
$ \langle \psi(\ve r,t)\psi(\ve 0,0)\rangle=(u_0^2\eta^2/2)\,
\exp\left[-|\ve r|^2/(2\eta^2)-|t|/\tau\right]$.
Green contours correspond to high vorticity of $\ve u(\ve r,t)$, blue contours to high strains.  Particle number
density: white (low density) to red (high density). Parameters:
$\ku=0.1$,$\st=10$,$t=165\tau$. {\bf b}, same but for
 $\ku =10$, $\st=0.025$, $t=0.32\tau$.
{\bf c} Parameter plane for inertial particles in mixing flows.
Region 1: clustering
is caused by ergodic \lq multiplicative amplification', see text.
Region 2: the non-ergodic \lq centrifuge' mechanism is important.
 Turbulent flows correspond to $\ku \sim 1$,
 strong clustering is  observed for $\st \sim 1$,
region 3. In region 4 the new perturbation
expansion is accurate (schematic).}  
\end{figure}
{\em Introduction}.
The dynamics of independent particles in complex mixing flows is
a problem of fundamental importance in the natural sciences, and in technology.
The motion of the particles is commonly approximated by
\begin{eqnarray}
\ddot{\ve r}=\gamma [\ve u(\ve{r},t)-\ve v]\,.
\label{eq:1}
\end{eqnarray}
Here $\ve r$ is the position
of a suspended particle, and ${\ve v}=\dot{\ve r}$ is its velocity.
Dots denote time derivatives,
$\gamma$ is the rate at which the inertial
motion is damped relative to the fluid, and ${\ve u}({\ve r},t)$ is
the velocity  of
a randomly mixing or turbulent incompressible flow.
It is a surprising fact that even though $\ve u$ is incompressible,
the suspended particles may nevertheless cluster together \cite{Max87}.
The effect is illustrated in Fig. \ref{fig:1}{\bf a},{\bf b}.
Possible consequences
of this phenomenon have been discussed in a wide
range of contexts: rain
initiation from turbulent clouds \cite{Fal02,Sha03,Wil06},
grain dynamics in circumstellar accretion disks \cite{Bra99,Wil08},
and plankton dynamics \cite{Rei03}, to name but a few.

Despite its significance, clustering of particles
in mixing flows is still not well understood.
Two very different explanations of the phenomenon
have been put forward. Maxey \cite{Max87} discussed the problem
in the limit of small inertia, corresponding
to small values of the \lq Stokes number' $\st=(\gamma\tau)^{-1}$.
Here $\tau$ is the relevant characteristic time scale of the flow
(the Kolmogorov time in turbulent flows, for example).
For $0 < {\rm St}\ll 1$, the particles are argued to be
\lq centrifuged' out of
regions of high vorticity of $\ve u(\ve r,t)$.
The approach rests on
instantaneous correlations between  particle positions
and fluid velocities. It
has been refined by many authors \cite{Pin97,Bal01,Chu05}
and is commonly used to interpret results of experiments
\cite{Loh08,Gib10}, and of direct numerical simulations \cite{Cen06}.
But the \lq centrifuge' mechanism relies on a small-${\rm St}$ expansion,
while clustering in turbulent flows is observed
to be strongest and thus of most interest at
$\st \sim 1$.

A very different clustering mechanism
was  proposed \cite{Dun05} in the limit of large $\st$
and small \lq Kubo numbers'. 
The Kubo number \cite{Dun05,vKa81} $\ku=u_0\tau/\eta$
characterises  fluctuations of $\ve u(\ve r,t)$
($u_0$ and $\eta$  being its characteristic velocity and length
scales).  In the limit $\st\rightarrow \infty$ 
(and $\ku\rightarrow 0$ so that $\epsilon^2 \equiv \ku^2 \st/2$ remains constant), 
the particles experience the velocity field
as a white-noise signal, and sample it in
an ergodic fashion: the fluctuations of $\ve u(\ve r_t,t)$
(and its derivatives) along 
a particle trajectory $\ve r_t$ are indistinguishable
from the fluctuations of $\ve u(\ve r_0,t)$
at the fixed position $\ve r_0$.
This case corresponds to region 1 in the phase diagram Fig. \ref{fig:1}{\bf c},
and in this limit the instantaneous
configuration of $\ve u(\ve r_t,t)$ is irrelevant to the
dynamics of the suspended particles. But they
may nevertheless cluster by the mechanism
of \lq multiplicative amplification':
small line-, area-, and volume elements randomly expand and contract.
Depending upon whether the random product of expansion and contraction
factors increases or decreases as $t\rightarrow \infty$,  one may observe
fractal clustering in region 1.
The fractal dimension $d_{\rm L}$ is
determined by the history of these factors.  It can be computed in terms of
\lq Lyapunov exponents' \cite{Dun05,Som93}.

Figs.~\ref{fig:1}{\bf a},{\bf b} show both mechanisms at work:
at large values of $\st$, (region 1 in Fig.~\ref{fig:1}{\bf c}) 
there is no discernible influence
of the instantaneous $\ve u(\ve r_t,t)$ on the particle distribution.
At small $\st$,  (region 2 in Fig.~\ref{fig:1}{\bf c}), 
by contrast, the particles are seen to avoid regions of high vorticity
(Fig.~\ref{fig:1}{\bf b}, similar to Fig.~8 in \cite{Vas06}).
In short, in limiting cases (regions 1 and 2 of Fig. \ref{fig:1}{\bf c})
the mechanisms of clustering are understood. But how ergodic and non-ergodic
effects compete in the major part of the phase diagram Fig. \ref{fig:1}{\bf c}
is not known
(in particular not for the experimentally most relevant region 3
where $\ku$ and $\st$ are of order unity).  
In \cite{Bec07} non-ergodic effects were characterised by correlating
the degree of clustering with the probability of particles avoiding
rotational regions of the flow. The interpretation of these numerical
results, however, is complicated by the fact that this probability
is significantly enhanced even when clustering is weak.
In order to understand the importance of non-ergodic and ergodic effects,
an analytical theory is required, valid at finite Stokes and Kubo numbers.

{\em Summary}. Here we derive a perturbation
expansion for the Lyapunov exponents of particles in
random flows, valid at finite $\st$ and $\ku$.
We compute the Lyapunov fractal dimension $d_{\rm L}$,
and characterise fractal clustering in terms of
the \lq dimension deficit' $\Delta_{\rm L} = d-d_{\rm L}$.
For particles suspended
in two-dimensional random Gaussian incompressible flows, 
the new theory yields reliable 
results for $\ku < 0.2$ for arbitrary values of $\st$, region
4 in Fig. \ref{fig:1}{\bf c} (schematic).
We find, first, that for small values of $\st$, 
the \lq centrifuge' mechanism dominates,
and $\Delta_{\rm L} = 6 \ku^2 \st^2$, consistent with \cite{Bec03,Wil07}.
Second, in the limit of $\ku\rightarrow 0$ at finite values of $\st$, 
non-ergodic
effects remain important. Third, in region 1 of Fig. \ref{fig:1}{\bf c},
clustering is found to be entirely
due to ergodic \lq multiplicative amplification' \cite{Dun05},
and $\Delta_{\rm L} = 12 \epsilon^2 \propto \st$.
Fourth, in general we find that the ergodic mechanism
contributes substantially to clustering, unless $\st$ is very small.
Fifth we show 
by numerical simulations of the model that
at $\ku\sim 1$ and $\st \sim 1$,
ergodic \lq multiplicative amplification' makes
a substantial contribution to the observed clustering.

{\em Method and results}. Eq.~(\ref{eq:1}) cannot be explicitly solved, 
since $\ve u$ depends upon the particle position at time $t$.
The implicit solution of Eq.~(\ref{eq:1}) becomes
(dimensionless variables $\ve r'=\ve r/\eta$, $t'=t/\tau$, $\ve v'=\ve v/u_0$,
 and $\ve u'=\ve u/u_0$)
\begin{eqnarray}
\label{eq:rsolution}
\delta \ve r_t &\equiv& \ve r_t\!-\!\ve r_0=\ku\,\Big[\st(1-{\rm e}^{-t/{\rm St}} )\ve v_0\\
&&
\hspace*{6mm}+{\st}^{-1}\int_{0}^{t} {\rm d}t_1\int_{0}^{t_1}{\rm d}t_2\,
{\rm e}^{-(t_1-t_2)/{\rm St}} \ve u(\ve r_{t_2},t_2)\Big]\,.
\nonumber
\end{eqnarray}
Here and in the following the primes are omitted. 
We seek an approximate solution by expanding $\ve u(\ve{r}_t,t)$ in powers of
$\delta\ve r_t$.
Since according to Eq.~(\ref{eq:rsolution}),  $\delta\ve r_t$  is of order $\ku$, iteration
generates an expansion of $\ve u(\ve r_t,t)$ in  powers of $\rm Ku$. To second order, for example, we find
\begin{widetext}
\begin{eqnarray}
&&u_\alpha(\ve r_t,t)=
u_\alpha(\ve r_0,t)
+\frac{\ku}{\st}\int_{0}^{t} {\rm d}t_1\int_{0}^{t_1}{\rm d}t_2e^{-(t_1-t_2)/\st}
\sum_\beta \frac{\partial u_\alpha}{\partial r_\beta}(\ve r_0,t)u_\beta(\ve r_0,t_2)
\cr&&
+\frac{\ku^2}{\st^2}\int_{0}^{t} {\rm d}t_1\int_{0}^{t_1}{\rm d}t_2\int_{0}^{t_2} {\rm d}t_3\int_{0}^{t_3}{\rm d}t_4e^{-(t_1-t_2+t_3-t_4)/\st}
\sum_{\beta,\delta}\frac{\partial u_\alpha}{\partial r_\beta}(\ve r_0,t)\frac{\partial u_\beta}{\partial r_\delta}(\ve r_0,t_2)u_\delta(\ve r_0,t_4)\cr
\cr&&
+\frac{1}{2}\frac{\ku^2}{\st^2}\int_{0}^{t} {\rm d}t_1\int_{0}^{t} {\rm d}t_2\int_{0}^{t_1}{\rm d}t_3\int_{0}^{t_2}{\rm d}t_4e^{-(t_1+t_2-t_3-t_4)/\st}
\sum_{\beta,\delta}\frac{\partial^2 u_\alpha}{\partial r_\beta\partial r_\delta}(\ve r_0,t)u_\beta(\ve r_0,t_3)u_\delta(\ve r_0,t_4)
+\Ordo(\ku^3)\,.
\label{eq:uexpansion_Ku2}
\end{eqnarray}
\end{widetext}
Here Greek indices denote the components of $\ve u$, and 
for our purposes $\ve v_0$ can be set to zero.
The coefficients in 
(\ref{eq:uexpansion_Ku2}) are expressed
in terms of $\ve u$ and its derivatives
at the fixed position $\ve r_0$,
with known statistical properties.
This procedure can in principle be extended to any order in $\ku$, but in practice
it is limited by the number of nested integrals appearing in (\ref{eq:uexpansion_Ku2}) 
 for higher orders. A C-program was written to symbolically
evaluate the integrals.
One may expand 
other functionals of the particle trajectories, such as the
strain matrix $\ma A(\ve r_t,t)$
with elements $A_{\alpha\beta}={\partial u_\alpha}/{\partial r_\beta}$.
Previous analytical results on the clustering of inertial particles  \cite {Meh04,Dun05} rest
on the \lq ergodic assumption'
that the distribution of the strain matrix $\ma A(\ve r_t,t)$
at the particle position $\ve r_t$ can be approximated by its distribution
at $\ve r_0$.
This is satisfied in region 1,  but what
are the corrections outside this region? 
For the two-dimensional incompressible  ($\tr \ma A=0$) random flow
described in Fig.~\ref{fig:1} we find:
\begin{eqnarray}
\overline{\tr \ma A^2} &\equiv & \Big\langle
\lim_{T\rightarrow\infty}\!\frac{1}{T}\!\!
\int_0^T\!\!\!\!{\rm d}t\, \tr \ma A^2(\ve r_t,t)\Big\rangle
= \frac{6 \ku^2\st}{(1\!+\!\st)^2(1\!+\!2{\rm St})}\nonumber\\
&&\hspace*{-1cm}-\frac{2{\rm Ku}^4 {\rm St} (4 + 52 {\rm St} + 293 {\rm St}^2 + 548 {\rm St}^3 + 297 {\rm St}^4)}{(1+{\rm St})^4(2+{\rm St})(1+2{\rm St})^2 (1+3{\rm St})}
\label{eq:av}
\end{eqnarray}
to order ${\rm Ku}^4$. 
The average in (\ref{eq:av})
consists of a long-time average along the particle trajectory $\ve r_t$,
and an average over initial conditions $\ve r_0$
(denoted by $\langle \cdots\rangle$).
At finite values of the Kubo number,
$\overline{\tr \ma A^2}$ differs
from its ergodic average (which vanishes in incompressible flows):
the dynamics is not strictly ergodic.
In compressible flows, we find
$\overline{\tr \ma A}\neq 0$ (the ergodic average still vanishes).
This is consistent with a result \cite{Wil09a} for
the average strain in the advective limit
of a one-dimensional (compressible) model.
It was shown in \cite{Wil09a} that the
average strain must be taken into account
to obtain the known result  \cite{Bal01}
for the advective Lyapunov exponent in this model.

The question is now: how does non-ergodicity affect the spatial
distribution of the particles?
The latter 
is characterised by the Lyapunov
exponents of the particle flow, obtained by linearising Eq.~(\ref{eq:1}).
The maximal exponent $\lambda_1$ is given by
\begin{eqnarray}
\label{eq:lyapunov_def}
\lambda_1
\!&=&\!
\ku \lim_{T\to\infty}\frac{1}{T}\int_0^{T}\!\!\!\!{\rm d}t\,\ve n_1(\ve r_t,t)\cdot\ma Z(\ve r_t,t)\ve n_1(\ve r_t,t)\,,\\
\dot{\ma Z}\!&=&\!{\st}^{-1}(\ma A\!-\!\ma Z)\!-\!\ku\, \ma Z^2,\,\dot{\ve n}_1\! =\!\ku\big(\ve n_2\cdot\ma Z\ve n_1\big)\ve n_2\,.
\label{eq:lan2}
\end{eqnarray}
Here $\ve  n_1$ ($\ve n_2$)
is the unit vector in the $\delta\ve r$ ($\dot{\delta \ve r}$)-direction, and
$\ma Z$ is the matrix with elements
$Z_{\alpha\beta} = {\partial v_\alpha}/{\partial r_\beta}$.
In region 1, Eqs.~(\ref{eq:lyapunov_def},\ref{eq:lan2})
were solved in \cite{Meh04,Dun05}.
At finite values of
$\ku$,  we compute $\lambda_1$ by 
generalising the procedure that led to Eq. (\ref{eq:uexpansion_Ku2}).
Starting from the implicit solution (\ref{eq:rsolution})  of (\ref{eq:1}), and the implicit
solutions of (\ref{eq:lan2}):
\begin{eqnarray}
\nonumber
\ma Z(\ve r_t,t)&=&{\rm e}^{-t/{\rm St}} \ma Z(\ve r_0,0)+
\int_0^t\!\!{\rm d}t_1
{\rm e}^{-\frac{t-t_1}{\rm St}}
\big[\ma A(\ve r_{t_1},t_1)/\st\nonumber\\
&&\hspace*{2.55cm}-\ku\,\ma Z(\ve r_{t_1},t_1)^2\big]\,,\\
\ve n_1(\ve r_t, t)&=&\ve n_1(\ve r_0,0)+\ku\!
 \int_0^t\!\!{\rm d}t_1[\ve n_2(\ve r_{t_1},t_1) \cdot\ma Z(\ve r_{t_1},t_1)\nonumber\\
&&\hspace*{1.5cm}\times\ve n_1(\ve r_{t_1},t_1)]\ve n_2(\ve r_{t_1},t_1)\,,
\end{eqnarray}
we expand $\ma A(\ve r_t,t)$, $\ma Z(\ve r_t,t)$, $\ve n_1(\ve r_t,t)$, and $\ve n_2(\ve r_t,t)$
in powers of $\delta \ve r_t$. Iterating and averaging along particle trajectories as
well as over initial conditions  yields an expansion of $\lambda_1$
in powers of $\ku$, with $\st$-dependent coefficients.
The sum $\lambda_1+\lambda_2=\ku\,\overline{\tr \ma Z}$
is computed in a similar fashion.
For particles
in a two-dimensional incompressible random Gaussian flow
(c.f. Fig.~\ref{fig:1})
we find
\begin{widetext}
\begin{eqnarray}
\lambda_1&=& {\ku^2} - {\ku^4}\frac{6+16\st+16\st^2+15\st^3+5\st^4}{(1+\st)^3}
+{\ku^6}\Big[\frac{1692+16464\st+68987\st^2+165269\st^3+258832\st^4}{6(1+\st)^5(2+\st)^2(1+2\st)^2}\nonumber\\
&&+\frac{301534\st^5+296820\st^6
+247404\st^7+153480\st^8+62136\st^9+14400\st^{10}+1440\st^{11}}{6(1+\st)^5(2+\st)^2(1+2\st)^2}\Big]\,, \label{eq:7}\\
\lambda_1+\lambda_2&=& -6{\ku^4}\frac{\st^2(1+3\st+\st^2)}{(1 +\st)^3}\label{eq:8}\\
&&+2{\ku^6}\st^2\frac{8+92\st+598\st^2+2509\st^3+5760\st^4+7176\st^5+5052\st^6+2076\st^7+480\st^8
+48\st^9}{(1+\st)^5 (2+\st)^2(1+2\st)^2}\,,\nonumber
\end{eqnarray}
\end{widetext}
to order $\ku^6$. This is our main result.
Eq.~(\ref{eq:7}) yields the ergodic expansion \cite{Meh04}
$\lambda_1/(\gamma\tau) = 2\epsilon^2 - 20 \epsilon^4 + 480\epsilon^6+\ldots $
in region 1.

As $\st\rightarrow 0$, Eq.~(\ref{eq:8}) reflects
Maxey's non-ergodic centrifuge mechanism:
According to  (\ref{eq:1}),
a particle is advected by an effective velocity field $\ve v$
with  compressibility ${\nabla\cdot\ve v}={\tr\ma Z}$.
Maxey's result \cite{Max87,Sha03} is obtained by
expanding $\ma Z\approx \ma Z^{(0)}+Z^{(1)}\st$ in Eq.~(\ref{eq:lan2}). One finds:
$\nabla\cdot\ve v=-\ku\,\st\left.{\tr\ma A^2}\right|_{\st=0}$ 
(note that $\ma A(\ve r_t,t)$ depends upon the Stokes number because
the particle trajectory $\ve r_t$ depends upon ${\rm St}$).
This result shows that particles
tend to aggregate in regions of high strain or low vorticity.
However, since the velocity field is homogeneous, this lowest-order
term averages to zero in incompressible flows. Expanding $\ma Z$ to second
order in ${\rm St}$, one finds
$\overline{\nabla\cdot\ve v}
=-\ku\,\st^2\left.{\partial_{\rm St}{\overline{\tr(\ma A^2)}}}\right|_{\st=0}$.
Inserting Eq.~(\ref{eq:av}) yields
$\lambda_1+\lambda_2 = \ku\,\overline{\ve \nabla \cdot\ve v}
= -6\ku^4\st^2 + 4 \ku^6\st^2+\Ordo({\rm Ku}^8)$,
consistent with the $\st\rightarrow 0$ limit of (\ref{eq:8}).

We conclude that our new expansion (\ref{eq:7},\ref{eq:8})
of the Lyapunov exponents correctly describes  the
different clustering mechanisms in the advective and
ergodic regions (Fig.~\ref{fig:1}{\bf c}):
 the \lq centrifuge' effect and ergodic \lq multiplicative amplification'.
More importantly,
Eqs.~(\ref{eq:7},\ref{eq:8}) allow to determine how
the relative importance of the two mechanisms depends on $\ku$ and $\st$, 
as follows.

\begin{figure}[t]
\centerline{
\includegraphics[width=8.5cm]{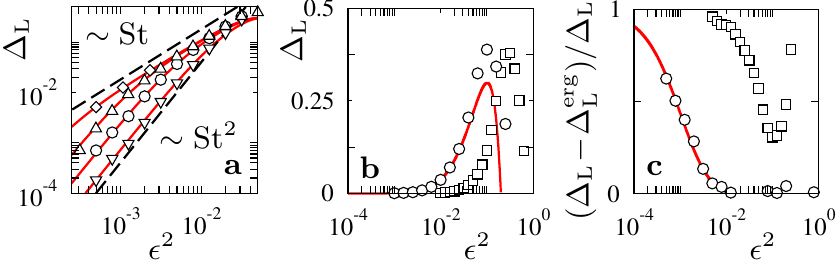}}
\caption{\label{fig:2} {\bf a} Fractal dimension deficit
as a function of $\epsilon^2 = \ku^2\st/2$. Numerical simulations
of the model described in Fig.~\ref{fig:1}
($\ku=0.02$ ($\Diamond$), $\ku=0.05$ ($\bigtriangleup$),
$\ku=0.1$ ($\circ$), and $\ku=0.2$ ($\bigtriangledown$);
theory according to Eqs.~(\ref{eq:7},\ref{eq:8}), solid lines, and limiting
behaviours $\Delta_{\rm L} \propto \st$ and $\Delta_{\rm L} \propto \st^2$ (dashed lines).
{\bf b}   Same but
for larger values of $\epsilon^2 =\ku^2\st/2$. Numerics, $\ku=0.1$ ($\circ$)
and $\ku=1$ ($\Box$); theory,  Eqs.~(\ref{eq:7},\ref{eq:8}),
$\ku=0.1$ (solid line).
{\bf c} Relative importance of non-ergodic and ergodic contributions (see text),
symbols and parameters as in {\bf b}.}
\end{figure}

In  two-dimensional incompressible flows,
the fractal dimension deficit is given by
$\Delta_{\rm L} = (\lambda_1+\lambda_2)/\lambda_2$.
Fig.~\ref{fig:2}{\bf a} shows that the new theory explains the limiting
cases in region 1 in the parameter plane ($\Delta_{\rm L} =
6 \,\ku^2\, {\rm St}=12\,\epsilon^2$), and in region 2
($\Delta_{\rm L} = 6\,\ku^2 \st^2 $).
The theory also explains
the cross-over between these two behaviours and compares well with
results of numerical simulations. In order to further increase
the accuracy, more terms than computed in Eqs.~(\ref{eq:7},\ref{eq:8})
must be included. The series is likely to be asymptotic requiring
re-summation, and there may be additional
non-analytic contributions \cite{Dun05,Meh04}.
We note that for a slightly different
estimate of the fractal dimension (the \lq correlation dimension'),
the lowest-order
behaviour of the dimension deficit in region 2,
$\Delta_{\rm C} \propto \st^2$, was computed
using different methods in \cite{Bal01,Chu05,Zai03,Fal04,Ijz10}.
For the model described in Fig.~\ref{fig:1}, we find
$\Delta_{\rm C} = 12\,\ku^2\st^2$
to lowest order in $\ku$ and in $\st$ in region 2,
and $\Delta_{\rm C} = 24\,\epsilon^2$ in region 1 (see also \cite{Wil10b}).

These results raise the question: how important are
non-ergodic contributions
to fractal clustering at larger values of $\st$
and at finite Kubo numbers? The answer
is summarised in Fig.~\ref{fig:2}{\bf b},{\bf c}
showing the dimension deficit
$\Delta_{\rm L}$ compared to an ergodic
approximation, $\Delta_L^{\rm erg}$.
The latter incorporates finite-time correlations of the velocity field $\ve u$ 
but neglects non-ergodic effects. Ergodic approximations to the 
Lyapunov exponents and to $\Delta_{\rm L}$ (referred to as 
$\lambda_{1,2}^{\rm erg}$
and $\Delta_{\rm L}^{\rm erg}$)
are obtained by expanding 
Eqs.~(\ref{eq:lyapunov_def},\ref{eq:lan2})
as before, but replacing 
$\ma A(\ve r_t,t)$ in (\ref{eq:lan2}) with $\ma A(\ve r_0,t)$. 
The resulting analytical expressions for  $\lambda_{1,2}^{\rm erg}$
are determined by the fluctuations of  $\ma A(\ve r_0,t)$. This is in 
contrast to $\lambda_{1,2}$, Eqs. (\ref{eq:7},\ref{eq:8}), which 
are determined by the fluctuations of $\ve u(\ve r_0,t)$ and its derivatives. 
The ergodic approximation allows
for finite values of $\ku$ and $\st$ but must fail in the limit $\st\rightarrow 0$,
since the \lq centrifuge' mechanism is not accounted for.
In particular, to lowest order in $\ku$
the exponents $\lambda_1^{\rm erg}$ and $\lambda_2^{\rm erg}$ 
are found to depend upon $\st$. The exact exponents (\ref{eq:7},\ref{eq:8})
by contrast, are independent of $\st$ 
to lowest order in $\ku$: $\lambda_1 = \ku^2$ and $\lambda_2 = -\ku^2$.
This implies in particular that earlier results for the Lyapunov
exponents in region 1 \cite{Meh04,Dun05} 
are in fact exact to lowest order in $\ku$ for arbitrary values of $\st$.
This is due to the cancellation of two errors: neglecting non-ergodic effects,
and neglecting finite-time correlations.

Non-ergodic effects dominate the clustering when
$(\Delta_{\rm L}-\Delta_{\rm L}^{\rm erg})/\Delta_{\rm L}$ is close to unity
(they are negligible when this ratio is close to zero).
We have determined $\lambda_{1,2}^{\rm erg}$ and $\Delta_{\rm L}^{\rm erg}$ 
to order 
$\ku^6$ \cite{Wil10}.  We have also performed computer simulations of this 
\lq ergodic model' by calculating monodromy matrices (Eq.~(30) in \cite{Wil07}) with $\ma A(\ve r_0,t)$ evaluated at the fixed position $ve r_0$.  
Fig.~\ref{fig:2}{\bf c}  shows that for $\ku=0.1$, 
non-ergodic effects dominate at small values of
$\epsilon^2$
but are negligible when clustering is largest,
near the peak in $\Delta_{\rm L}$ shown in Fig.~\ref{fig:2}{\bf b}.
Also shown are the analytical result for $\Delta_{\rm L}$
derived from Eqs.~(\ref{eq:7},\ref{eq:8}),
and the corresponding expansion of
$(\Delta_{\rm L}-\Delta_{\rm L}^{\rm erg})/\Delta_{\rm L}$.
We observe good agreement.  For ${\rm Ku}=1$, by contrast,
 the first terms in the perturbation expansion
(\ref{eq:7},\ref{eq:8}) do not give reliable results. But
computer simulations show that non-ergodic effects
are present for the whole range of Stokes numbers
displayed in Fig.~\ref{fig:2}{\bf c}. However,
Fig.~\ref{fig:2}{\bf c} also clearly shows that
both mechanisms contribute in region 3.  
We emphasise that ergodic clustering by \lq multiplicative amplification'
makes a substantial contribution in this region,
$(\Delta_{\rm L}-\Delta_{\rm L}^{\rm erg})/\Delta_{\rm L} \approx 0.3$.


\begin{thebibliography}{}
\bibitem[Maxey (1987)]{Max87}
M. R. Maxey, {J.~Fluid Mech.} {\bf 174} (1987) 441

\bibitem{Sha03}R. A. Shaw, Annu. Rev. Fluid. Mech. {\bf 35} (2003) 183.

\bibitem{Fal02}  G. Falkovich, A. Fouxon and G. Stepanov,
Nature {\bf 419} (2002) 151-154.

\bibitem{Wil06} M. Wilkinson, B. Mehlig and V. Bezuglyy,
Phys. Rev. Lett.  {\bf 97} (2006) 048501.

\bibitem{Bra99} A. Bracco {\em et al.},
Phys. Fluids {\bf 11} (1999) 2280

\bibitem[Wilkinson, Mehlig and Uski, (2008)]{Wil08}
M. Wilkinson, B. Mehlig and V. Uski,
Astrophys. J. Suppl. Ser. {\bf 176} (2008) 484

\bibitem{Rei03} R. Reigada {\em et al.},
Proc. R. Soc. Lond. B {\bf 270} (2003) 875

\bibitem{Pin97} M. Pinsky and A. Khain,
Q. J. Met. Soc. {\bf 123} (1997) 165

\bibitem{Bal01} E. Balkovsky, G. Falkovich, and A. Fouxon,
Phys. Rev. Lett. {\bf 81} (2001)  2790

\bibitem{Chu05} J. Chun {\em et al.},
J. Fluid Mech. {\bf 536} (2006) 219

\bibitem{Loh08}
E. Calzavarini {\em et al.}, Phys. Fluids {\bf 20} (2008) 040702

\bibitem{Gib10} M. Gibert, H. Xu, and E. Bodenschatz,
arxiv:1002.3755

\bibitem{Cen06}
M. Cencini {\em et al.}, J. Turbulence {\bf 7} (2006) 1

\bibitem{Dun05}
K. P. Duncan {\em et al.}, Phys. Rev. Lett. {\bf 95} (2005) 240602.

\bibitem[van Kampen (1981)]{vKa81} N. G. van Kampen,
{\em Stochastic processes in physics and chemistry}, 2nd ed.,
North-Holland, Amsterdam 1981.

\bibitem{Som93} J. Sommerer and  E. Ott, Science {\bf 259} (1993) 351

\bibitem{Vas06} L. Chen, S. Goto, and J. C. Vassilicos, J. Fluid Mech. {\bf 553} (2006) 143

\bibitem{Bec07} J. Bec {\em et al.}, Phys. Rev. Lett. {\bf 98} (2007) 084502  

\bibitem{Meh04} B. Mehlig and M. Wilkinson,
Phys. Rev. Lett. {\bf 92} (2004) 250602

\bibitem{Bec03} J. Bec, Phys. Fluids {\bf 15} (2003) L81


\bibitem{Wil07} M. Wilkinson {\em et al.}, Phys. Fluids {\bf 19} (2007) 113303

\bibitem[Wilkinson (2009)]{Wil09a} M. Wilkinson, arXiv:0911.2917

\bibitem{Zai03} L. I. Zaichik and V. M. Alipchenkov,
Phys. Fluids {\bf 15} (2003) 1776

\bibitem{Ijz10} R. H. A. Ijzermans, E. Meneguz, and M. W. Reeks,
J. Fluid Mech. {\bf 653} (2010) 99

\bibitem{Fal04} G. Falkovich and Pumir, Phys. Fluids {\bf 16} (2004) L47

\bibitem{Wil10b} M. Wilkinson, B. Mehlig, and K. Gustavsson,
Europhys. Lett. {\bf 89} (2010), 50002

\bibitem{Wil10} K. Gustavsson and B. Mehlig, unpublished.
We have obtained the ergodic series also in a different way:
by  generalising the one-dimensional approach described by 
M. Wilkinson,  J. Stat. Phys. {\bf 139} (2010) 345.


\end{thebibliography}
\end{document}